\documentclass[10pt,onecolumn]{article}
 \pdfoutput=1
\usepackage{amsfonts,amssymb,amsmath}
\usepackage{graphicx}
\usepackage{balance}
\usepackage[font=small]{caption}
\usepackage{paralist}
\usepackage{array}
\usepackage{setspace}
\usepackage{fancyhdr}
\usepackage{caption}
\usepackage{multicol}
\usepackage{geometry}
\usepackage{changepage}
\usepackage[super,numbers,sort&compress]{natbib}

\newcolumntype{x}[1]{%
>{\centering\hspace{0pt}}p{#1}}

\def\ni{\noindent}

\def\be{\begin{eqnarray}}
\def\ee{\end{eqnarray}}

\newcommand\AppendixFigureWidth{0.8333\textwidth}

\begin{document}
\sloppy

\twocolumn[ 
\begin{center}
\textbf{\Large Evaluation of dynamic causal modelling and Bayesian model selection using simulations of networks of spiking neurons}\\
\bigskip
Matthew G. Thomas\\
Anaesthetic Department, Bedford Hospital, United Kingdom\\
Former affiliation: School of Medicine and Biomedical Sciences, University of Oxford, United Kingdom\\
matthew.thomas1@jesus.oxon.org
\end{center}
\ni Inferring the mechanisms underlying physiological and pathological processes in the brain from recorded electrical activity is challenging. Bayesian model selection and dynamic causal modelling aim to identify likely biophysical models to explain data and to fit the model parameters. Here, we use data generated by simulations to investigate the effectiveness of Bayesian model selection and dynamic causal modelling when applied at steady state in the frequency domain to identify and fit Jansen-Rit models. We first investigate the impact of the necessary assumption of linearity on the dynamics of the Jansen-Rit model. We then apply dynamic causal modelling and Bayesian model selection to data generated from simulations of linear neural mass models, non-linear neural mass models, and networks of discrete spiking neurons. Action potentials are a characteristic feature of neuronal dynamics but have not previously been explicitly included in simulations used to test Bayesian model selection or dynamic causal modelling. We find that the assumption of linearity abolishes the qualitative transitions seen as a function of the connectivity parameter in the original Jansen-Rit model. As with previous work, we find that the recovery procedures are effective when applied to data from linear Jansen-Rit neural mass models, however, when applying them to non-linear neural mass models and networks of discrete spiking neurons we find that their effectiveness is significantly reduced, suggesting caution is required when applying these methods.
\bigskip] 

\begin{onehalfspace}


\bigskip
\ni\textbf{\large{Introduction}}\\[1mm]
\ni Electroencephalography (EEG) provides a non-invasive method for spatially resolved measurement of the electrical activity of the brain\cite{Buzsaki12}. EEG is a valuable clinical and research tool, however, it remains challenging to infer key properties of the underlying network, such as channelopathies, from EEG recordings. 

Dynamic causal modelling\cite{Friston03,Penny04,David06,Moran08a,Kiebel09,Kahan13,Friston19a,Jafarian19} (DCM) aims to infer the parameters of biophysical models from recorded electrical activity. Biophysical models are constructed such that model parameters have a direct interpretation in terms of network properties, for example, connection weights or ion channel dynamics. By inferring the parameters of biophysical models from EEG or local field potential (LFP) recordings, DCM may provide a means to infer properties of the underlying neuronal network. While DCM seeks to infer the parameters of a particular biophysical model, choosing a particular model requires knowledge of the structure of the underlying network. Bayesian model selection (BMS)\cite{Penny04,Moran11a,Friston19a} seeks to infer which of a range of models best explains the recorded data.

To date, validation of these methods has taken two forms. Firstly, the techniques have been applied to data generated from models with the same structure as those underlying the DCM and BMS techniques and the recovered parameters were compared to those used as input to the original model \cite{Friston03,Penny04,Moran08a,Moran08,Papadopoulou15,Friston23}. The second approach to validation has been the application to data from human subjects and animals\cite{Friston03,David06,Moran08a,Moran08,Moran11,Moran11a,Moran13,Moran14,Symmonds18a,vanWijk18}. 

The use of simulations in validation has the advantage of exact knowledge of the network structure and parameters producing the synthetic data, allowing direct comparison with those recovered. To date, however, this approach has been applied by generating data with equations of exactly the form employed in the recovery methods\cite{Friston03,Penny04,Moran08a,Moran08,Papadopoulou15,Friston23}. This is likely to have overestimated their effectiveness. Furthermore, the model equations that have been used to generate synthetic data have had no explicit spiking dynamics and thereby omit a characteristic feature of the dynamics of networks of neurons. 

Studies employing data from humans and animals have been of two types. Some studies used patients with previously diagnosed channelopathies \cite{Gilbert16,Symmonds18a}. The strength of this approach is that the underlying pathology is known and thus can be used to validate the results of the analysis methods. The weakness, however, is the extremely limited number of patients and channelopathies. The alternative approach is to study evoked responses or heterogeneous disease groups. Much larger sample numbers can be obtained, however, the true network properties are not completely certain, making validation of the results difficult \cite{Friston03,David06,Moran08a,Moran08,Moran11,Moran11a,Moran13,Moran14,Symmonds18a,vanWijk18}. 

Here, we further investigate the validity of DCM and BMS, considering their implementation based on Jansen-Rit models in frequency space and applied at steady state\cite{Moran08a}. We start by analysing the effects of the assumptions used to generate the necessary transfer functions. We then apply DCM and BMS to simulated data. We move from simulations of linear neural mass models, through non-linear neural mass models, to simulations of networks of discrete neurons with spiking dynamics. This work builds on past computational studies, with their control and exact knowledge of the network structure and parameters to be recovered, by adding spiking activity, a key feature of neuronal function not implemented in previous forward models and a key difference between the underlying system of interest and the neural mass models that DCM and BMS fit to data. 

The use of transfer functions to fit models to data in frequency space requires an assumption of linearity\cite{Moran08a}. We find that implementing this linearity in Jansen-Rit models causes a qualitative change in the model dynamics, losing the transitions otherwise seen as a function of coupling parameters. As in previous work\cite{Moran08a}, we find good agreement between the models and parameters recovered from data generated from linearised Jansen-Rit models, however, we find that the introduction of non-linearity in the neural mass models and networks of discrete neurons significantly reduces the effectiveness of both DCM and BMS.

\begin{figure*}
\centering
\includegraphics[width= 0.7\textwidth]{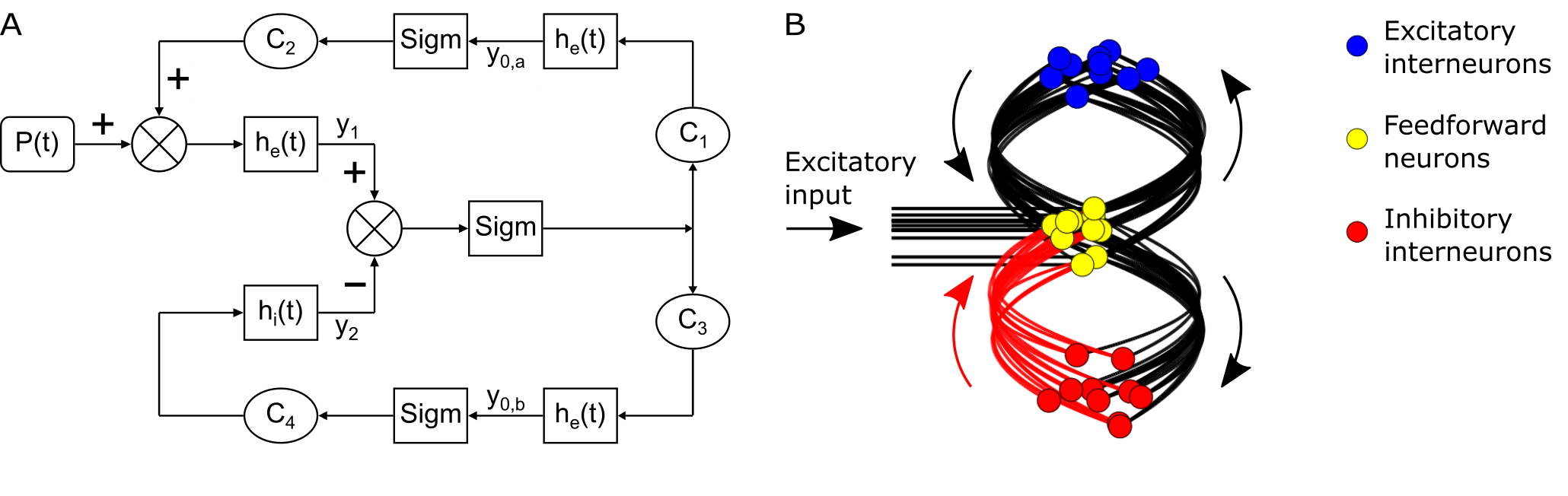}
\caption{Model structures. (A) The Jansen-Rit neural mass model, composed of populations of feedforward pyramidal cells, excitatory interneurons, and inhibitory interneurons. $P(t)$ is the external input; $h_e(t)$ and $h_i(t)$ are the excitatory and inhibitory impulse response functions respectively; $Sigm$ are sigmoid functions; coupling constants are as defined in the text; and addition and subtraction symbols indicate excitation and inhibition respectively. (B) Network of spiking neurons. Connections reflect the structure of the Jansen-Rit model. Black lines represent excitatory connections, red lines represent inhibitory connections. The directions of the connections are indicated by arrows. A reduced number of neurons in each population is shown for clarity.}
\label{fig:Figure1}  
\end{figure*}

\bigskip
\ni\textbf{\large{Jansen-Rit Neural Mass Models}}\\[1mm]
\ni Neural mass models simplify the complexity of networks of neurons to more tractable averaged quantities\cite{Wilson72,Wilson73}. The Jansen-Rit model\cite{Jansen93,Jansen95} is a model of a local cortical circuit. It consists of components reflecting three populations of neurons: one population of feedforward pyramidal neurons and two populations of interneurons, one excitatory and one inhibitory, that form local feedback loops (Figure~\ref{fig:Figure1}A). The feedforward pyramidal neurons receive external input and give external output. 

Each population in the model can be thought of as being composed of two blocks. The first block takes a mean input firing rate and outputs a mean membrane potential. The impulse response of this block has the form of a post-synaptic potential, for $t \geq 0$: $aAte^{-at}$ for an excitatory input spike and: $bBte^{-bt}$ for an inhibitory input spike, where $A$ and $B$ are amplitude parameters and $a$ and $b$ are reciprocals of the time constants of the post-synaptic potential. The second block takes the mean membrane potential from the first block and produces a mean output spike rate. This transformation takes the form of a sigmoid: 
\begin{equation}
S(v) = \frac{v_{Max}}{1 + e^{r(v_0-v)}} - \frac{v_{Max}}{1 + e^{rv_0}}
\end{equation}
where $r$ controls the gradient of the sigmoid function, $v_{Max}$ reflects the maximum firing rate of the population, and $v_0$ reflects the potential at which the firing rate is half of maximum. The second term in the expression shifts the function so that $S(0) = 0$.

The external input, $P(t)$, is white noise. Here it is uniformly distributed across the range from 120 to 320 pulses per second. The parameters $C_1$, $C_2$, $C_3$, and $C_4$ control the strength of coupling between the populations and the variables $y_{0,a}$, $y_{0,b}$, $y_{1}$, and $y_{2}$ correspond to the average membrane potentials within the blocks, as in Figure~\ref{fig:Figure1}A. The output from the model is the average potential of the feedforward pyramidal neurons: $y_1 - y_2$. Illustrative parameter values are presented in Table~\ref{Tb:JRParameters}. 

\begin{center}
\begin{tabular}{ c | c }
\hline			
Parameter & Example value\\
\hline
$a$ & $100$ s$^{-1}$\\
$A$ &  3.25 mV\\
$b$ & $50$  s$^{-1}$\\
$B$ & 22 mV\\
$C_1$ & 300\\
$C_2$ & $0.8 C_1$\\
$C_3$ & $0.25 C_1$\\
$C_4$ & $0.25 C_1$\\
$v_{Max}$ & $5$ s$^{-1}$\\
$v_0$ & $6$  mV\\
$r$ & $0.56$ mV$^{-1}$\\
\hline  
\end{tabular}
\captionof{table}{Illustrative Jansen-Rit model parameters.}
\label{Tb:JRParameters}
\end{center}

\ni The model can be described by four coupled second order ordinary differential equations:

\begin{equation}
\frac{d^2y_{0,a}(t)}{dt^2} = a A C_1 S(y_1(t) - y_2(t))-2 a \frac{dy_{0,a}(t)}{dt} - a^2 y_{0,a}(t)
\label{eq:JR1}
\end{equation}
\begin{equation}
\frac{d^2y_{0,b}(t)}{dt^2} = a A C_3 S(y_1(t) - y_2(t))-2 a \frac{dy_{0,b}(t)}{dt} - a^2 y_{0,b}(t)
\label{eq:JR2}
\end{equation}
\begin{equation}
\frac{d^2y_{1}(t)}{dt^2} = a A (P(t) + C_2 S(y_{0,a}(t))-2 a \frac{dy_{1}(t)}{dt} - a^2 y_{1}(t)
\label{eq:JR3}
\end{equation}
\begin{equation}
\frac{d^2y_{2}(t)}{dt^2} = b B C_4 S(y_{0,b})-2 b \frac{dy_{2}(t)}{dt} - b^2 y_{2}(t)
\label{eq:JR4}
\end{equation}

\begin{figure*}
\centering
\includegraphics[width= 0.8333\textwidth]{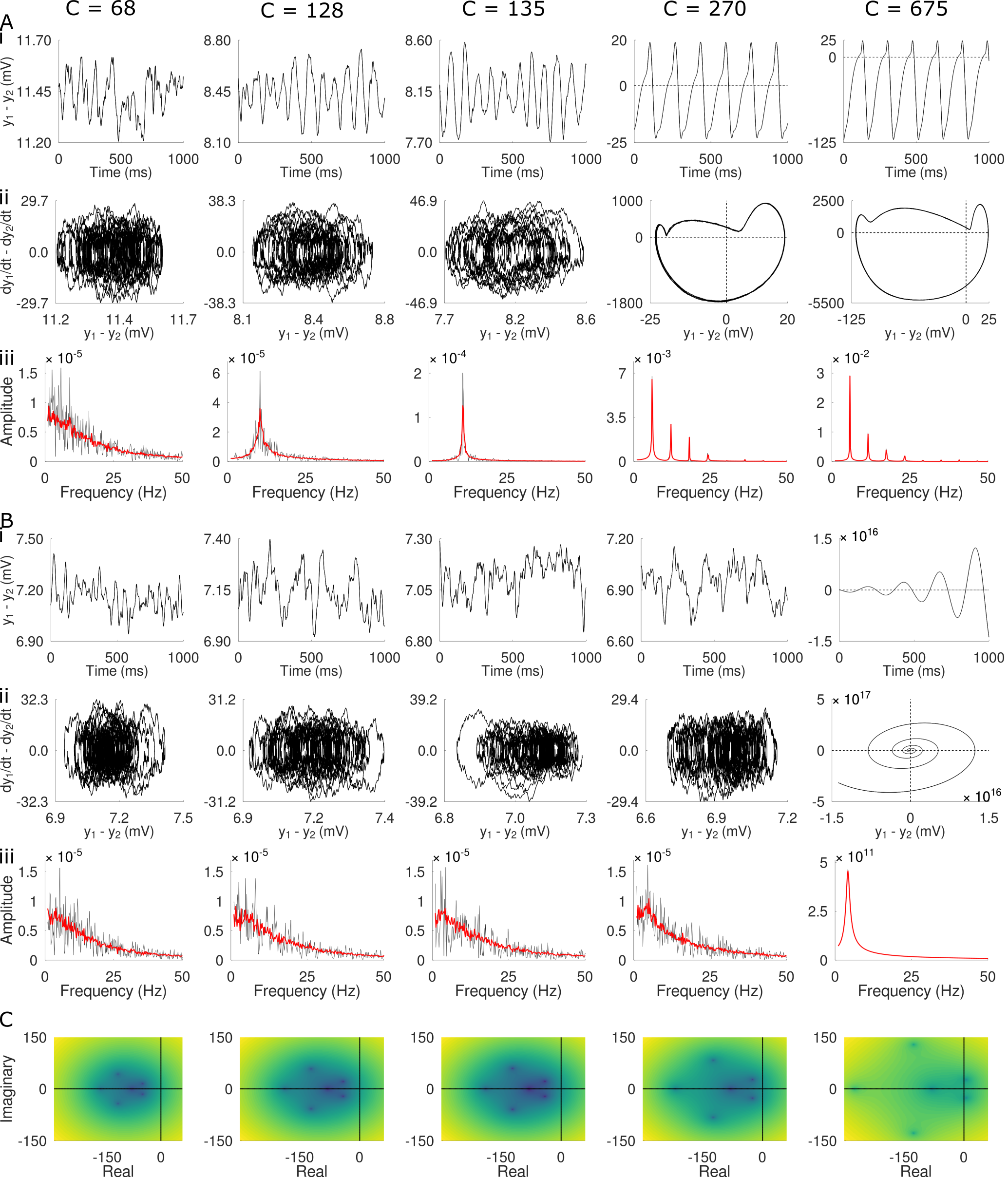}
\caption{Dynamics of the Jansen-Rit model. (A) Dynamics with the original sigmoid relationship between the average membrane potential and output firing rate. (B) Dynamics with the sigmoid replaced with a linearised relationship. The relevant coupling constant is shown at the top of each column. (i) Time courses of $y_1-y_2$ at steady state. (ii) Phase portraits: $\dot{y}_1-\dot{y}_2$ plotted against $y_1-y_2$. (iii) Amplitudes of the Fourier transforms of the time courses shown in (i) (black) and the average of 20 time courses (red). (C) Stability analysis. The poles of the transfer functions of the linearised models (B) are shown in dark blue. Poles with positive real values are associated with unstable dynamics.}
\label{fig:Figure2}  
\end{figure*}

\ni The dynamics of the Jansen-Rit model have been well studied\cite{Jansen93,Jansen95,Ableidinger17,Ahmadizadeh18}. The model produces a range of dynamics depending on the connectivity parameter, $C$, where $C_1 = C$. As the connectivity parameter is increased there is a transition from noisy output to periodic oscillations (Figure~\ref{fig:Figure2}A). 

\begin{figure}
\includegraphics[width= \columnwidth]{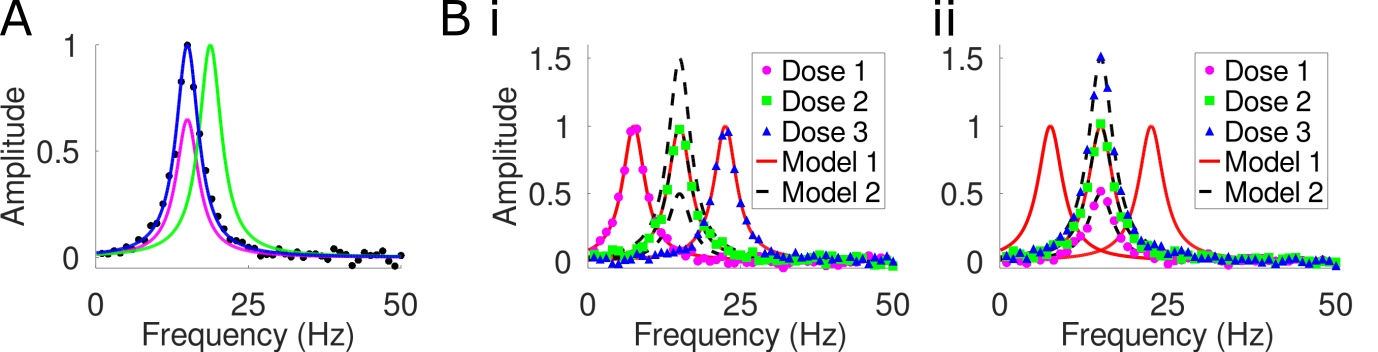}
\caption{Illustration of DCM and BMS. (A) DCM seeks to identify the most likely set of parameters to explain recorded data. The black circles show illustrative data points. The three coloured curves show the same function with different parameter sets. The blue curve corresponds to the parameter set most likely to explain the data. (B) BMS seeks to identify the model that best explains measured data sets. The coloured points show illustrative data sets. The coloured curves show different models. (i) Model 1 best explains the data sets. (ii) Model 2 best explaines the data sets.}
\label{fig:Figure3}  
\end{figure}

\bigskip
\ni\textbf{\large{Dynamic Causal Modelling}}\\[1mm]
\ni DCM aims to infer probability distributions for the parameters of biophysical models using macroscopic observations\cite{Friston03} (Figure~\ref{fig:Figure3}A); thereby allowing inference of the mechanistic basis of observed phenomena at the microscopic level from non-invasive measurements\cite{Friston03,Moran08a}. DCM has been implemented in a number of forms, with varying models and recording modalities \cite{David03,Friston03,Penny04,David06,Moran08,Moran08a,Stephan08,Kiebel09,Moran11a,Kahan13,vanWijk18,Jafarian19,Friston19a}. 

Here we will consider the implementation used in Moran \textit{et al}\cite{Moran08a} as it is conceptually simple and so readily allows identification of its strengths and limitations. This approach uses a Jansen-Rit model as the biophysical model whose parameters are to be inferred. The data considered are electrophysiological recordings, such as EEG or LFP time series. Probability distributions for the parameters of the model are inferred from the data in frequency space. The transformation to frequency space for the observed data is readily obtained by taking the Fourier transform of the time series. The analytic transformation to the form for the model in frequency space requires the equations of the model to be linear. To achieve this, the sigmoid functions in the Jansen-Rit model are approximated by linear functions, with proportionality constant $\gamma$: 
\begin{equation}
S(v) = \frac{v_{Max} r e^{r v_0}}{(1 + e^{r v_0})^2} v = \gamma v
\end{equation}
With this assumption, the transfer function of the model can be found by taking the Laplace transform of the equations. The transfer function for the Jansen-Rit model equations in the previous section is readily derived:
\begin{equation}
H(s) = \left( \frac{(s+a)^2}{aA} - \frac{aAC_1C_2\gamma^2}{(s+a)^2}+\frac{bBC_3C_4\gamma^2}{(s+b)^2}\right)^{-1}
\label{eq:TF}
\end{equation}
In frequency space the model can be thought of as a filter operating on the spectra of the input, to produce the spectra of the measured output. We can write this as: $Y(s) = H(s)U(s)$, where $Y(s)$ and $U(s)$ are the spectra of the model output and input respectively and $H(s)$ is the transfer function for the system.

We wish to infer the parameters of the transfer function, and, therefore, the Jansen-Rit model, from the Fourier transform of data. There are a number of ways to achieve this. A Bayesian approach can be used to include prior beliefs about parameter values and variational approaches, such as an expectation-maximization scheme, can be used to invert the model\cite{Moran08a}.

In our study we wish to treat the simplest cases to identify the effects of key assumptions on the effectiveness of the DCM approach. To this end we will consider white noise input to the model, yielding, on average over many instances: $\langle |Y(s)| \rangle \propto \langle |H(s)| \rangle$. We will also assume no prior knowledge of the parameters and calculate their likelihood functions by exhaustive enumeration over a wide range around the true model parameters.

Assuming information about the phase of the input is unknown, we consider only the amplitude of the input and output spectra. Taking the amplitude amounts to combining two components with independent, normally distributed, noise in quadrature. The result is that the output spectra is the transfer function multiplied by noise with a Rayleigh distribution. 

The likelihood of a given model parameter set, $\theta$, is calculated given the spectra, $W(f)$, of an experimental recording as follows. We write $x(f)$ as the ratio:
\begin{equation}
x(f) = \frac{|W(f)|}{|H(f,\theta)|}
\end{equation}
We consider the probability density function of the Rayleigh distribution, with scale parameter $\sigma$:
\begin{equation}
R(x,\sigma) = \frac{x}{\sigma^2}e^{\frac{-x^2}{2\sigma^2}}
\end{equation}
We can then calculate the likelihood of a given parameter set given observed data with samples at discrete frequencies:
\begin{equation}
L(\theta \vert W) = \prod_{i} R(x(f_i))
\end{equation}
We note that the Rayleigh distribution has a single parameter, $\sigma$, and that the mean of the distribution can be expressed: $\sigma\sqrt{\frac{\pi}{2}}$. For data generated by the Jansen-Rit model, if we allow ourselves knowledge of the amplitude of the input noise we can utilise the formula for the standard deviation of uniformly distributed white noise ($\frac{1}{\sqrt{12}}(P_{Max}-P_{Min})$) to divide $W(f)$  by this amplitude. The likelihoods can then be calculated with $\sigma = \frac{1}{\sqrt{2}}$ in the Rayleigh distribution. If we allow ourselves to assume the input noise is white, but do not allow knowledge of the amplitude, we calculate the likelihood by normalising both the observed spectra and the transfer function by dividing by their means. In this case the likelihoods are calculated with $\sigma = \sqrt{\frac{2}{\pi}}$. A similar approach can be employed for models of discrete spiking neurons, where treating the input spikes as a Poisson distributed train of delta functions allows calculation of the scale parameter for the resulting Rayleigh distributed noise in frequency space, if the input frequency is known, using the formula $\langle U(s) \rangle \propto \frac{\sqrt{(\pi \langle P \rangle)}}{2}$. In each case an appropriate correction is required to account for the convention used with regard to the number of samples in the numerical Fourier transform of the model output.

\newpage
\ni\textbf{\large{Bayesian Model Selection}}\\[1mm]
\ni BMS aims to infer the most likely model to explain a set of observations (Figure~\ref{fig:Figure3} B). For example, we can imagine administering a drug that has some effect on the brain, while not knowing the mechanism of action of the drug. We consider different possibilities, for example it may prolong the duration of excitatory postsynaptic potentials (EPSPs) or it may reduce the amplitude of inhibitory postsynaptic potentials (IPSPs) (for simplicity we will imagine a drug that has a single effect). We imagine having EEG recordings associated with different doses of drug. Bayesian model selection then infers the most likely model to explain the data.
\ni Given models $M_i$ and data $D$, Bayes theorem gives:
\begin{equation}
P(M_i \vert D) = \frac{P(D \vert M_i) P(M_i)}{P(D)}
\label{eq:}
\end{equation}
Here we will consider the simple case of constant priors, yielding as a result the maximum likelihood estimate (MLE): 
\begin{equation}
P(M_i \vert D) \propto P(D \vert M_i)
\label{eq:}
\end{equation}
For each model we integrate over all possible parameters sets, $\theta$:
\begin{equation}
P(D \vert M_i) = \int P(D \vert M_i, \theta) P(\theta \vert M_i) d\theta
\label{eq:}
\end{equation}
Assuming $P(\theta \vert M_i)$ is a constant independent of $M_i$, this becomes: 
\begin{equation}
P(M_i \vert D) \propto \int P(D \vert M_i, \theta) d\theta
\label{eq:}
\end{equation}
Here we will consider the simple case where our set of trial models is restricted to 11 models, for each of which a single parameter, $d$, acts linearly on one of the parameters in the transfer function. For example, the transfer function in the case where the action of the intervention increases the amplitude of EPSPs, $A$, is: 
\begin{equation}
M(s, d) = \left( \frac{(s+a)^2}{adA} - \frac{adAC_1C_2\gamma^2}{(s+a)^2}+\frac{bBC_3C_4\gamma^2}{(s+b)^2}\right)^{-1}
\label{eq:BMSTF}
\end{equation}
When calculating the likelihood of each model we treat noise as for DCM. We can then calculate the relative likelihood of the different models and identify the one most likely to explain the data set.


\bigskip
\ni\textbf{\large{Materials and Methods}}\\[1mm]
\ni Simulations and data analysis were carried out using bespoke scripts in Octave-6.2.0. Simulations were carried out with time steps of 0.1 ms. Systems were initialised with firing rates and potentials of zero. For each simulation the time to steady state was established by comparing the Fourier transform of a shifting 1 s window of the output either to the model transfer function, in the case where the transfer function was exact for the model, or the final second of the simulation. The initial period of each simulation affected by transients was discarded. Simulations were repeated 20 times for each parameter set.


\bigskip
\ni\textbf{\large{Results}}\\[1mm]
\ni \textbf{Effect of linearisation on Jansen-Rit model dynamics}

\ni Deriving the transfer function for a Jansen-Rit model requires the sigmoid functions to be linearised. Working with the transfer function is then equivalent to inferring the parameters to a modified form of the Jansen-Rit equations. We start by investigating the effect of this linearisation on the dynamics of the Jansen-Rit neural mass model.

In the original Jansen-Rit model the sigmoid function bounds the firing rate between $\frac{-v_{Max}}{1+e^{r v_0}}$ and $v_{Max}\left(1-\frac{1}{1+e^{r v_0}}\right)$. Linearisation of the sigmoidal response removes these bounds on the firing rate. This results in a significant change in the model dynamics, as shown in Figure~\ref{fig:Figure2}B. Noisy dynamics persist for a larger range of values of the connectivity parameter. The dynamics eventually transition to a growing oscillation. This transition can be understood in terms of the poles of the transfer function (Figure~\ref{fig:Figure2}C). For small enough values of the connectivity parameter all poles of the transfer function have negative real components, corresponding to decaying dynamics. As the parameter is increased, poles cross the imaginary axis of the complex plane so that they have positive real components. This then corresponds to an unstable system where the output increases with time. The analytic form for the value of the connectivity parameter at which this transition occurs for this model is derived in the Appendix. For the parameter set used here, the transition occurs at approximately $C=606.6$.  

\medskip
\ni \textbf{Recovery of parameters from a linearised Jansen-Rit model}

\ni We wish to use DCM to infer the parameters of neural networks. We start by looking at the simplest case - applying DCM to data generated from a linearised Jansen-Rit model, such that the transfer function we use is exactly that for the model equations used to generate the data. We allow ourselves exact knowledge of the structure of the network, the form of the dynamics of the model, and all but one of the model parameters in turn. We then use DCM to calculate a maximum likelihood estimate for the unknown parameters. 

\begin{figure}
\includegraphics[width= \columnwidth]{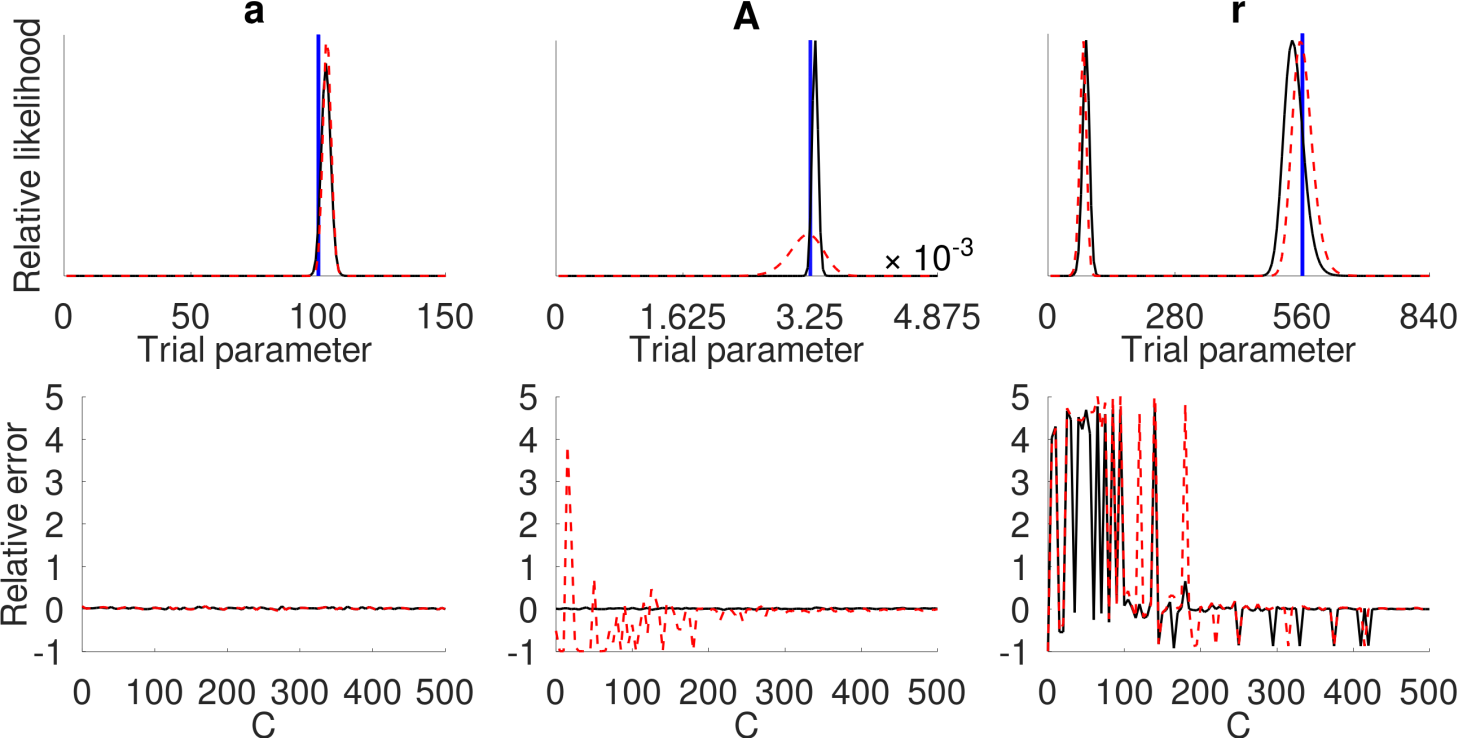}
\caption{Recovery of the parameters of the linearised Jansen-Rit model using DCM. Top row: Relative likelihood of parameter values (C = 250). The vertical blue bars indicate the value of the parameter used in the model. Bottom row: Relative error in the maximum likelihood estimates of the parameters as a function of the coupling constant, $C$, used in the model.
The black curves show the results when the input spike rate is known. The dashed red curves show the likelihoods derived when the input spike rate is unknown.}
\label{fig:Figure4}  
\end{figure}

Twenty time series were generated from the model at each of a range of connectivity constants, $C$, noting the dependence of the other connection constants on $C$. The likelihood function was then calculated for each parameter in turn, while the remaining parameters were set to their true values. We note that the likelihood function did not have to be calculated for every parameter as some parameters only appear in the transfer function (Equation~\ref{eq:TF}) as a product with other particular parameters. For example, $C_1$ and $C_2$ only appear in the transfer function as a product and so a relative change in one cannot be distinguished from a relative change in the other. In contrast, the constant $A$ appears in the product with $C_1$ and $C_2$, but also appears in other terms in the transfer function and so variation in $A$ can have effects distinguishable from variation in $C_1$ and $C_2$. 

The likelihood functions were calculated for each parameter both with and without allowing knowledge of the input rate to the model. Figure~\ref{fig:Figure4} shows both the likelihood functions and the relative error in the MLE for illustrative parameters. The exhaustive results are presented in the Appendix (Figures~\ref{fig:Figure9Appendix}, \ref{fig:Figure10Appendix}, \ref{fig:Figure13Appendix}, and \ref{fig:Figure14Appendix}). The likelihood function for most of the parameters was unimodal, however, it was bi-modal for $r$, with similar amplitudes at the two modes. As $C$ was varied, the MLE would usually relate to the true mode, but would sometimes be the second value (Figure~\ref{fig:Figure4}C). 

For most parameters, the error in the MLE was greater at smaller values of the connectivity parameter, $C$. This can be understood in terms of the smaller effect on the dynamics of the system for most parameters when the coupling is weak. The parameter $a$ could be accurately recovered regardless of the strength of the connection because it appears in a term of the transfer function even when $C_1 = 0$. The parameter, $A$, also appears in the transfer function when $C_1 = 0$ and so can be recovered with weak coupling when the amplitude of the input to the model is known. However, when the input to the model is unknown and the likelihood estimate is calculated with normalised spectra the effect of $A$ is lost and its value cannot be recovered for small values of the connectivity parameter.

\medskip
\ni \textbf{Recovery of parameters from a Jansen-Rit model without linearisation}

\ni Neuronal dynamics are inherently non-linear. We are interested in the effectiveness of DCM at recovering the parameters of such non-linear systems when the inference procedure inherently employs a linear model. 

In moving towards testing DCM on data generated from a spiking neural network we next apply DCM to time series generated from the original, non-linear, Jansen-Rit model. 

We apply the same approach as in the previous section, but replacing the linearised sigmoid with the original sigmoid function when generating the data. As we saw above, exchanging the sigmoid function with a linearised form has a significant impact on the dynamics of the model. The transfer function corresponds to the linear function while re-introduction of the sigmoid function to the Jansen-Rit model generates different dynamics with different amplitudes and peaks in its frequency spectra (Figure~\ref{fig:Figure2}A). As the value of $C_1$ is increased, the dynamics of the model deviate away from dynamics exhibited by the linearised model (Figure~\ref{fig:Figure2}B). The MLEs resulting from the application of DCM therefore become significantly less accurate than those calculated for the linearised model (Figure~\ref{fig:Figure5}). The exhaustive results are presented in the Appendix (Figures~\ref{fig:Figure11Appendix}, \ref{fig:Figure12Appendix}, \ref{fig:Figure13Appendix}, and \ref{fig:Figure14Appendix}). For the same reasons as in the linearised case, when the connectivity constant is small, all but $a$ and $A$ weakly influence the transfer function. Again, the likelihood function for $r$ is sometimes multimodal.

\begin{figure}
\includegraphics[width= \columnwidth]{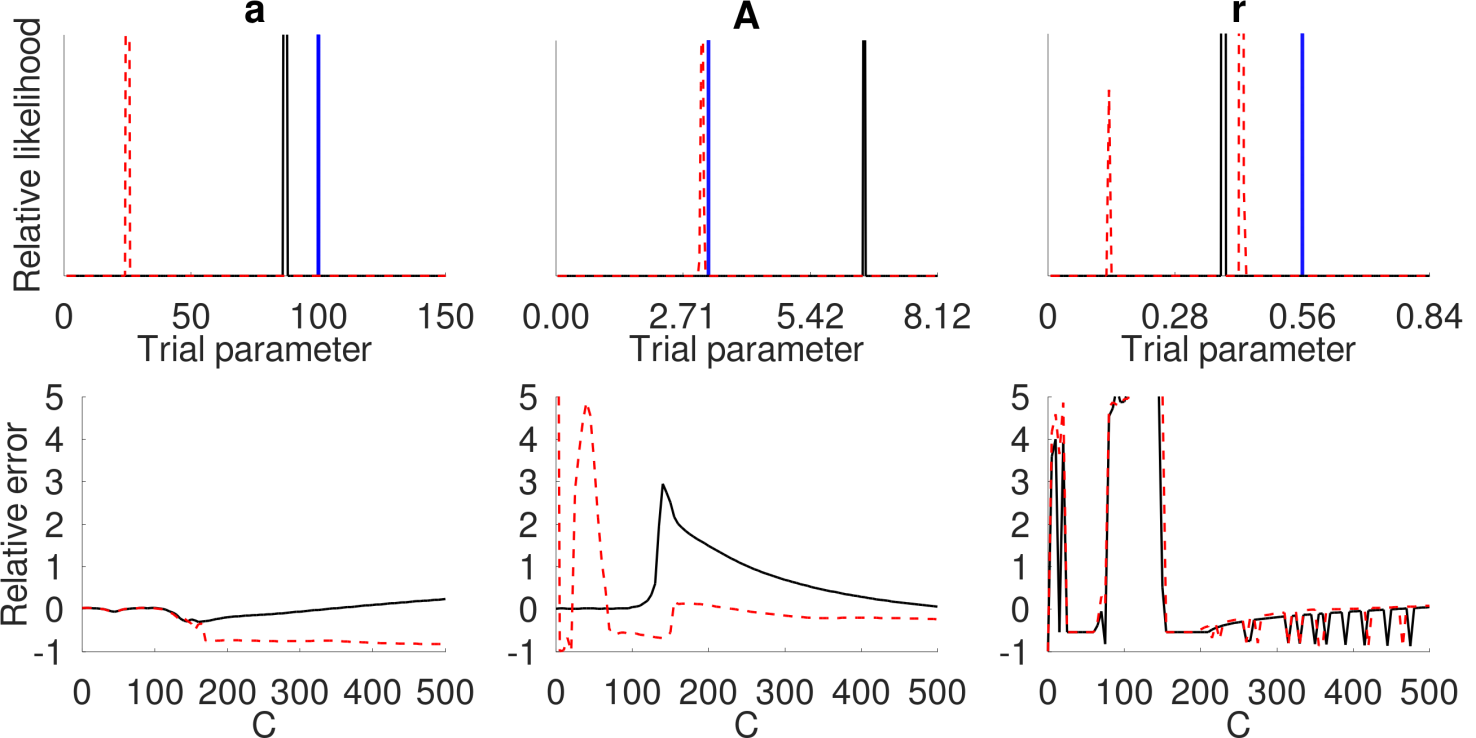}
\caption{Recovery of the parameters of the non-linear Jansen-Rit model using DCM. Top row: Relative likelihood of parameter values (C = 250). The vertical blue bars indicate the true value of the parameter used in the model. Bottom row: Relative error in the maximum likelihood estimates of the parameters as a function of the coupling constant, $C$, used in the model.
The black curves show the results when the input spike rate is known. The dashed red curve shows the likelihood derived from the same fit but with both the transfer function and Fourier transform normalised, reflecting the scenario where the input spike rate is unknown. Large scale turning points in the error, for example at $C\sim140$, correspond to transitions in the dynamics of the Jansen-Rit model. Smaller scale oscillations in the error, for example the error in the recovery of $r$ when $C>200$, correspond to shifts between the two modes of the relative likelihood.}
\label{fig:Figure5}  
\end{figure}

\newpage
\ni \textbf{Recovery of parameters from unconnected spiking neurons}

\ni Action potentials are a key characteristic of neuronal dynamics. Neural mass models consider averages over these events, yielding smoothed dynamics. Here we apply DCM to the effective LFP from a simulated population of unconnected spiking neurons.

To isolate the effect of spiking dynamics on the effectiveness of DCM we chose the subthreshold dynamics of our model neurons to match the form of the post-synaptic potential used in the Jansen-Rit model, with impulse response: $aAte^{-at}$. The only difference between the dynamics is then the spiking events. As there is no coupling, we can see from the equation for the sub-threshold dynamics that there are only two parameters to be recovered: $a$ and $A$.

\begin{figure}
\includegraphics[width= \columnwidth]{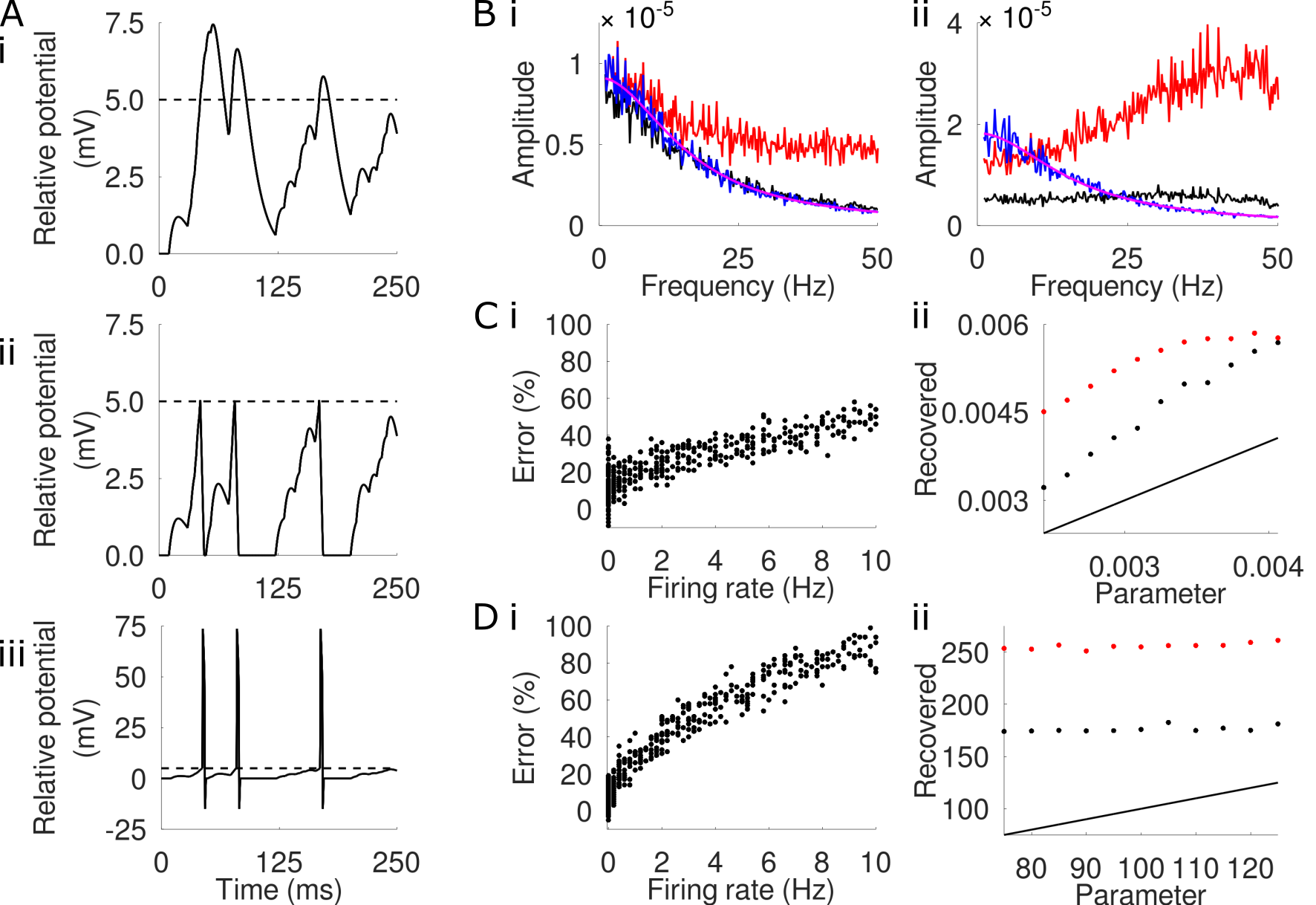}
\caption{Recovery of the parameters from a population of identical uncoupled spiking neurons using DCM. (A) Illustrative neuronal dynamics without spiking (i), with resets at spikes (ii), and with FitzHugh–Nagumo model action potentials (iii). (B) Amplitude of the Fourier transforms of the average potentials of populations of uncoupled neurons without spiking (blue, 100 neurons in each population, average of 20 populations shown), with resets at spikes (black), and with action potentials (red). The magenta line shows the transfer function of the equivalent non-spiking system. The average input rate to each neuron is 50 Hz (i) and 200 Hz (ii). (C) Error in the recovered EPSP amplitude for uncoupled neurons with resets at spikes, as a function of average firing rate (i) and EPSP amplitude used in the model (ii). (D) Error in the recovered EPSP time constant for uncoupled neurons with resets at spikes, as a function of average firing rate (i) and EPSP time constant used in the model (ii). The average input rates to the neurons in Cii and Dii are 100 Hz (black) and 200 Hz (red). The solid black lines indicate the true parameter values.}
\label{fig:Figure6}  
\end{figure}

We simulated three models: (i) neurons with an infinite threshold, giving purely subthreshold membrane potential dynamics (Figure~\ref{fig:Figure6}Ai); (ii) neurons with a finite threshold potential (5 mV above their resting potential) such that if it is reached, the neuronal potential returns to its resting potential linearly over a 4 ms refractory period (Figure~\ref{fig:Figure6}Aii); (iii) neurons behaving as in (ii) except that when the threshold is reached the potential follows a FitzHugh-Nagumo\cite{FitzHugh61} form action potential ($\tau = 12.5$, $a = 0.7$, $b = 0.8$) before returning to its resting potential (Figure~\ref{fig:Figure6}Aiii). 

For each model we simulated 20 instances, each composed of 100 neurons. The input to each neuron in each instance was a train of Poisson distributed spikes. For each instance, we generated 100 spike trains, one for each neuron. For a given instance we used the same set of input trains for each of the three models. For each instance of each model the effective LFP was generated by averaging the potential from 100 model neurons. While investigating the effect of spiking dynamics on the accuracy of DCM we allow ourselves knowledge of the average input rate to the neurons.

The first model produces an LFP that is identical to the dynamics of a Jansen-Rit model in which the connection parameters are all set to zero. The transfer function calculated above fits the spectra exactly (Figure~\ref{fig:Figure6}B) and the MLE accurately recovers the model parameters.

Subthreshold neuronal dynamics are known to dominate the dynamics of the LFP\cite{Buzsaki12}. The second model is, therefore, the most realistic reflection of the LFP from spiking neurons. As expected, the resets associated with firing events modify the frequency spectra, causing them to deviate from the transfer function (Figure~\ref{fig:Figure6}B). The error in the MLE of both parameters increases with firing rate (Figure~\ref{fig:Figure6}C and D). For the same input rate, the neuronal firing rate increases as the EPSP amplitude ($A$) is increased and as the reciprocal of the time constants ($a$) is decreased. As the parameters of the model are varied, the error in the recovered parameters vary with their impact on the neuronal firing rates (Figure~\ref{fig:Figure6}C and D). 

The third model is presented for completeness. Including action potentials in the calculated LFP causes its spectra to deviate substantially from the model transfer function (Figure~\ref{fig:Figure6}B). Action potentials may contribute to what is measured using a local field potential electrode, although the subthreshold dynamics are thought to dominate\cite{Buzsaki12}.

\newpage
\ni \textbf{Recovery of parameters from networks of spiking neurons}

\ni Finally, we investigated the effectiveness of DCM at recovering the parameters from a simulated network of coupled spiking neurons. We arranged the network to be as similar as possible to that of the Jansen-Rit model, both in terms of the structure of the network (Figure~\ref{fig:Figure1}B) and the inter-spike dynamics of the neurons. We only included the feedforward population and the inhibitory interneurons ($C_1 = C_2 = 0$) as this arrangement results in negative feedback, yielding a stable simulation. 

We simulated networks made up of 750 feedforward pyramidal neurons and 750 inhibitory interneurons. There were no connections within populations. There were connections between some feedforward neurons and some inhibitory neurons. Connections were introduced at random. Before modification of parameters, the connection probability between neurons of both populations was 10\% . 

The inter-spike dynamics were identical to the Jansen-Rit model response functions. As in model (ii) in the previous section, neurons had a finite threshold potential such that if it was reached, the neuronal potential returned to its resting potential linearly over a 1 ms refractory period \cite{Harrison15}. 

We wished to identify thresholds so that the average firing rates of the neurons in our model matched those of the populations of the Jansen-Rit model when the input firing rate to the model matched that of the Jansen-Rit model at steady state. On this basis, the thresholds of the feedforward neurons and the inhibitory interneurons were set as 13.52~mV and 4.96~mV respectively. The external input to the feedforward neurons was a Poisson distributed spike train with an average rate of 220~Hz.

\begin{figure*}
\centering
\includegraphics[width= 0.6666\textwidth]{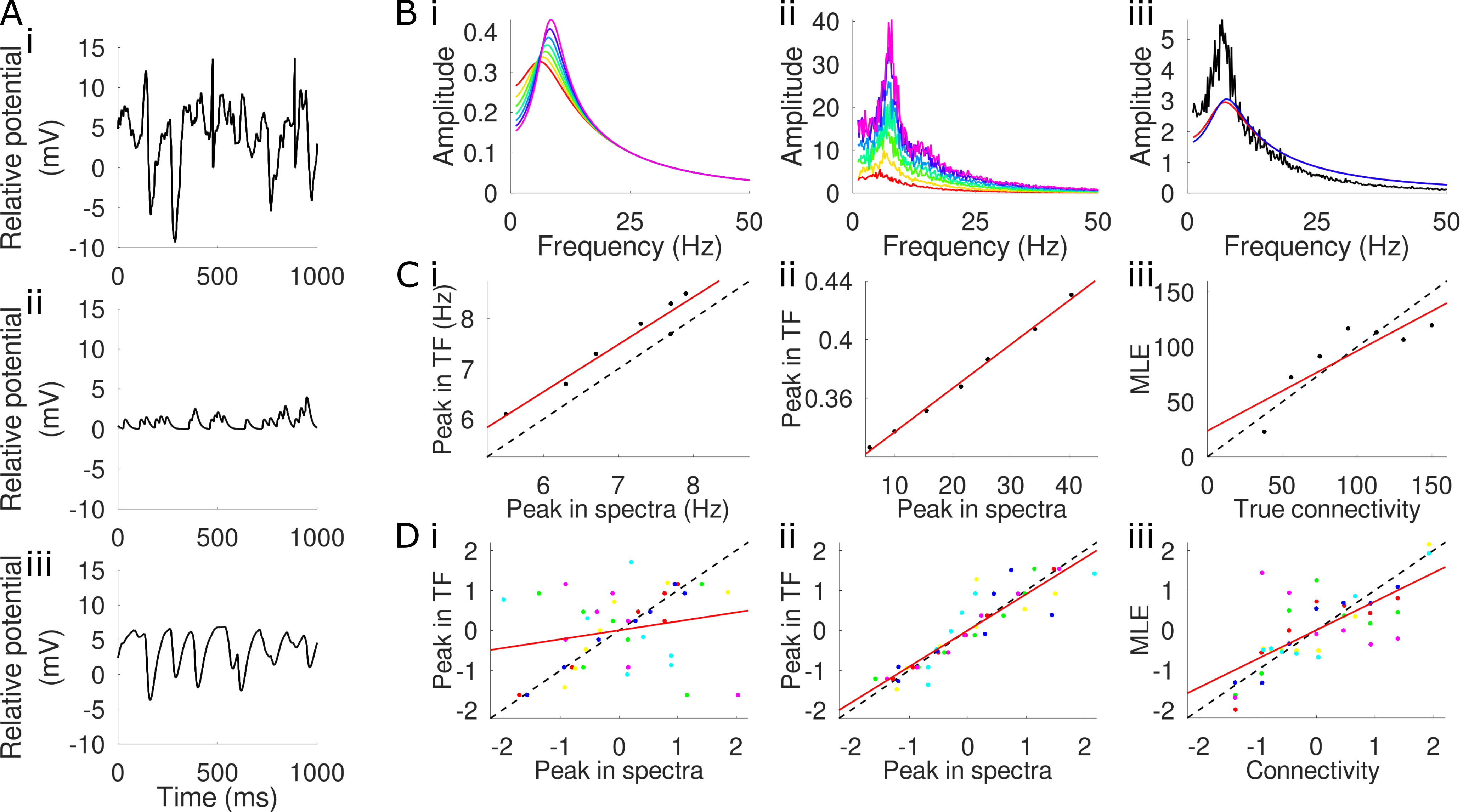}
\caption{Recovery of the parameters of a network of coupled spiking neurons using DCM. (A) Example membrane potential of a feedforward pyramidal neuron (i), of an inhibitory neuron (ii), and example LFP from a population of pyramidal neurons (iii). (B) The transfer function (i) and the spectra of the LFP of the spiking network (ii) as $C_3$, is varied from 50\% (red) to 200\% (magenta) of its standard value. (iii) The normalised transfer function (red) and spectra of the LFP (black) overlaid ($C_3 = 75$). The blue curve shows the transfer function for the maximum likelihood estimate of $C_3$ ($C_3 = 91$). (C) (i) The relationship between the frequency at the maxima of the transfer function and the spectra of the LFP as $C_3$ is varied. (ii) The relationship between the amplitude of the maxima of the transfer function and the spectra of the LFP as $C_3$ is varied. (iii) The relationship between the maximum likelihood estimate of the connectivity from fitting the transfer function to the spectra and the true value of $C_3$. (D) (i) The relative frequency at the maxima of the transfer function and the spectra of the LFP as each of the parameters is varied (colours reflect different parameters - full results in Appendix Figure~\ref{fig:Figure15Appendix}). (ii) The relationship between the relative amplitude of the maxima of the transfer function and the spectra of the LFP as each of the parameters is varied. (iii) The relationship between the relative maximum likelihood estimates of the parameters from fitting the transfer function to the spectra and the true parameters. In each case the solid red line shows a least squares fit to the data and the dashed black line shows the line $y=x$.}
\label{fig:Figure7}  
\end{figure*}

Networks were simulated for ten seconds. The average potential of the feedforward neurons was used as an effective LFP\cite{Buzsaki12} (Figure~\ref{fig:Figure7}A). The Fourier transform of the final five seconds of the effective LFP was used as the spectrum from each simulation. Networks were simulated 20 times for each parameter set. The model parameters ($a$, $A$, $b$, $B$, $C_3$, and $C_4$) were varied individually from 50\% to 200\% of the values in Table~\ref{Tb:JRParameters}, while the others were set as in Table~\ref{Tb:JRParameters}.

We first compared the spectra of the model LFPs to the transfer functions generated with the same parameters as used in the models (Figure~\ref{fig:Figure7}B). The form of the spectra is similar to that of the transfer function, typically exhibiting a single maximum, but does not match the shape exactly (Figure~\ref{fig:Figure7}B iii). We show illustrative results for the relationship between the spectra and transfer function as $C_3$ is varied in Figure~\ref{fig:Figure7}C and exhaustive results in the Appendix (Figure~\ref{fig:Figure15Appendix}). We quantify the relationship using Spearmans rank correlation coefficient (Table~\ref{Tb:Spearman}). As the parameters are varied the frequency at which the peak occurs in the spectra is correlated with the frequency at which the peak occurs in the transfer function for $A$, $b$, and $C_3$. The amplitude of the peak in the spectra is correlated with the amplitude of the peak in the transfer function for all parameters. The MLE of the parameters is correlated with the true value of the parameters for $A$ and $C_3$.

\begin{center}
\begin{tabular}{ c | c | c | c }
\hline			
Parameter & Frequency & Peak & MLE \\
\hline
$a$ & -0.45 & 1.00 & -0.32 \\
$A$ & 0.96 & 0.89 & 0.93 \\
$b$ & 0.96 & 0.89 & 0.39 \\
$B$ & -0.66 & 1.00 & -0.18 \\
$C_3$ & 0.94 & 1.00 & 0.86 \\
$C_4$ & -0.09  & 1.00 & 0.43 \\
\hline  
\end{tabular}
\captionof{table}{Spearmans rank correlation coefficients for the relationship between features of the model spectra and the equivalent transfer function as each parameter is varied.}
\label{Tb:Spearman}
\end{center}

\medskip
\ni \textbf{Bayesian Model Selection}

\ni The effectiveness of BMS was investigated by simulating models and varying each of the parameters in turn. The model parameters ($a$, $A$, $b$, $B$, $C_3$, and $C_4$) were varied individually from 50\% to 200\% of the values in Table~\ref{Tb:JRParameters}, while the others were set as in Table~\ref{Tb:JRParameters}. Each combination of parameters was simulated 20 times and the Fourier transform calculated as described above for DCM. BMS was then applied to the data sets for each parameter. 

\begin{figure}
\centering
\includegraphics[width= \columnwidth]{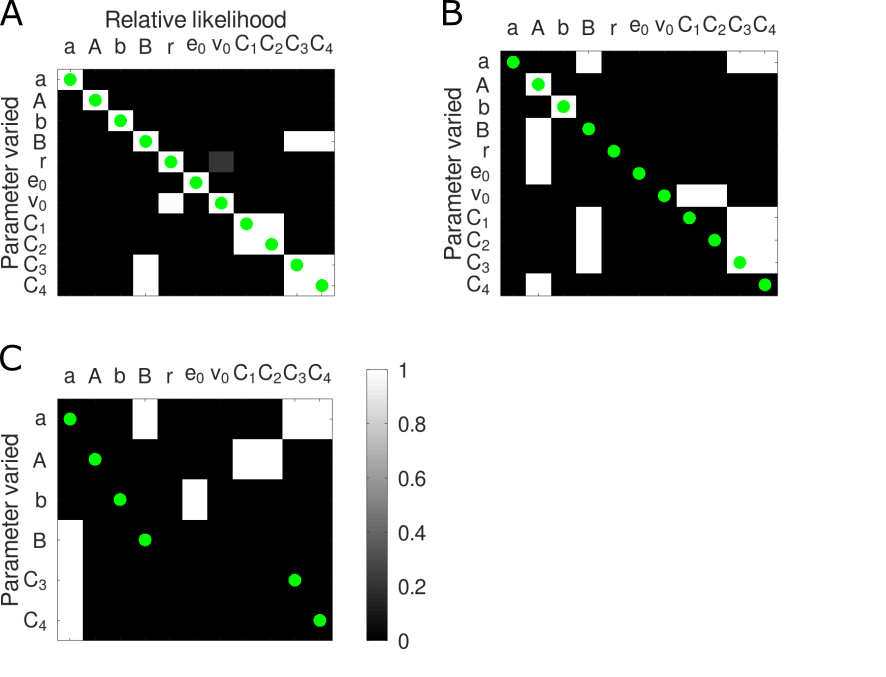}
\caption{Bayesian Model Selection. The rows correspond to the model parameter that was varied. Within each row the grayscale colour map reflects the relative likelihood of the simulated data being explained by variation in the parameter labelling the column. The green circles indicate the elements that correspond to the true parameter explaining the variation. (A) Data generated from the linearised Jansen-Rit model. (B) Data generated from the Jansen-Rit model without linearisation. (C) The effective LFP from the simulation of a network of spiking neurons.}
\label{fig:Figure8}  
\end{figure}

Application of BMS to data generated from the linearised Jansen-Rit model correctly identified the model parameter that was varied as one of the most likely to explain the data in every case. As can be seen from the form of the transfer function (Equation~\ref{eq:TF}) variation in a number of parameters have equivalent effects (for example: $B$, $C_3$, and $C_4$) and so cannot be distinguished by BMS and it identifies the members of the groups of parameters as equally likely (Figure~\ref{fig:Figure8}A).

Applying the same approach to data generated from the Jansen-Rit model without linearisation and to the model of a network of spiking neurons as described above found BMS to be less effective (Figure~\ref{fig:Figure8}B and C), not identifying any of the parameters correctly in the case of the model of the network of spiking neurons.

BMS was not applied to the data generated by the model of unconnected spiking neurons because the model only requires two parameters (EPSP time constant and amplitude) to describe its dynamics. 

\bigskip
\ni\textbf{\large{Discussion}}\\[1mm]
\ni DCM and BMS have the potential to infer the mechanistic basis of key physiological and pathological processes from recordings of averaged electrical activity, such as LFP or EEG. We have used simulations to investigate the effectiveness of DCM and BMS when applied at steady state, in frequency space, using a transfer function representation of a Jansen-Rit model. We started by investigating the effect of the linearisation required to determine a transfer function on the dynamics of the equivalent Jansen-Rit model. We found that making the assumption of linearity significantly changes the dynamics of the model. Applying DCM to simulated data from a linearised model works well when the transfer function is chosen to match the equations of the linearised model. We still find, however, that the likelihood function is sometimes multimodal. Some parameters only occur in the transfer function as a product with particular other parameters and so cannot be resolved individually. As the coupling between populations is reduced, the accuracy of the recovery of the parameters of the interneuron populations reduces. Knowledge of the input firing rate improves the accuracy of the recovery of the model parameters. Applying DCM to data generated from the original, non-linear, Jansen-Rit model was less accurate. As with the linear model, reduced coupling limited the recovery of the parameters of the interneuron populations. The parameters of a population of uncoupled neurons could be recovered from a simulated LFP in the absence of spiking dynamics, however, the introduction of spikes impaired the parameter recovery using DCM. The error in the recovery of the parameters increased with the frequency of spikes. The simulated LFP from a network of coupled spiking neurons, with a network structure reflecting that of the Jansen-Rit model, had a spectrum similar to that of the transfer function. As parameters in the model were varied in isolation the variation in the peak amplitude of the spectra was correlated with that of the transfer function. The frequency at which the peak occurred was correlated for some, but not all, of the parameters. There was a similar result for the MLE. Similar to the results for DCM, we found that BMS could accurately identify which parameter was varied for the linearised Jansen-Rit model. As follows from the form of the transfer function for the model, parameters that always appear together as a product could not be distinguished. The accuracy was less good for data generated from the Jansen-Rit model without linearisation and from a network of spiking neurons.

Any new methodology requires validation against a standard. Inferring the properties of the neuronal network that gives rise to an LFP or EEG recording is challenging. It is difficult to validate results as the properties of the true network often cannot be confirmed using alternative methods. Modelling studies have limitations, but allow knowledge of the underling parameters to evaluate against. Our study investigated the effectiveness of DCM and BMS applied at steady state, in the frequency domain, using a transfer function representation. As with previous work, we found DCM and BMS to be effective when applied to data generated from a linearised Jansen-Rit model. This is to be expected as, in this form, the model and transfer function are equivalent. Nonetheless, we note that there are still limitations in this simplified setting with this particular model as there are groups of parameters that only appear as a product in the transfer function and so their individual effects cannot be distinguished. With models that deviate from that relating exactly to the transfer function, either by incorporating non-linearity into the response of a Jansen-Rit model or by modelling distinct neurons with spiking dynamics, we found that DCM and BMS in this form was much less effective. DCM would be most accurate when applied to networks with a low firing rate and with as near linear firing rate in response to mean potential as possible. We found that changes could be identified better than absolute values.

The ability to infer the properties of a network of neurons from a recording of averaged electrical activity, such as EEG, would be incredibly valuable both for fundamental neuroscience and clinical investigations. For example, allowing the identification of a channelopathy causing epilepsy via non-invasive EEG recording, thereby allowing specific targeting of anti-epileptic drugs. Our investigation, although limited, found DCM and BMS in the form studied here to not be effective in achieving this aim. This suggests caution is required when employing these methods.

Our study was limited in a number of ways. Firstly, as with any modelling-based study, the synthetic data that we generated had fundamental differences from true EEG and LFP recordings as our models are extremely simplified in the dynamics they include. This is, however, not likely to lead to an underestimate of the effectiveness of the methods as the models used were specifically formulated to be as similar as possible to the Jansen-Rit model from which the relevant transfer function was generated. Our approach was handicapped by employing uninformative flat priors, thereby calculating maximum likelihood estimates. More informative priors would be likely to improve accuracy.

Our investigation is specific to the application of DCM and BMS at steady state, in frequency space, using a transfer function representation of a Jansen-Rit model. There are other similar approaches and it would be valuable to investigate their validity using similar modelling studies. It would also be valuable to test DCM and BMS on models of networks of spiking neurons with more realistic connectivity and conduction delays between neurons.


\newpage
\ni\textbf{\large{Code availability}}\\
Code to implement the models presented here can be found in the following GitHub repository: https://github.com/mgthomas1/Thomas2023


\bigskip
\ni\textbf{\large{Acknowledgements}}\\
I thank Prof. A. Farmery and Prof. R. McShane for their interest and support in developing this project.



\bibliographystyle{unsrtnat}
\bibliography{Bblgrphy}


\pagebreak
\onecolumn

\begin{adjustwidth}{1.5cm}{1.5cm}
\ni\textbf{\large{Appendix}}\\[1mm]
\ni \textbf{Stability analysis}

\ni Here we analyse the stability of the linearized Jansen-Rit model. The transfer function of the Jansen-Rit model considered is:
\begin{equation}
H(s) = \left( \frac{(s+a)^2}{aA} - \frac{aAC_1C_2\gamma^2}{(s+a)^2}+\frac{bBC_3C_4\gamma^2}{(s+b)^2}\right)^{-1}
\end{equation}

\ni If the real parts of all of the poles of the transfer function are negative the model is stable, if the real part of any of the poles are positive the system is unstable. While keeping the other parameters constant, we seek the value of $C$ that results in a transition from all poles having a real part that is negative to having a pole (or pair of poles) with a real part that is zero.

We seek $C$ such that there is an $s = 0 + il$ where the denominator of the transfer function is zero while the numerator is finite. This yields a pole with zero real part. For the denominator to be zero both the real and imaginary components must be zero.

\ni For the real part of the denominator to be zero:
\begin{equation}
a^2 - l^2 - \frac{a^2A^2C_1C_2\gamma^2(a^2-l^2)}{(a^2-l^2)^2+(2al)^2}+\frac{aAbBC_3C_4\gamma^2(b^2-l^2)}{(b^2-l^2)^2+(2bl)^2}=0
\label{eq:RlPrtDnmntr}
\end{equation}
\ni For the imaginary part of the denominator to be zero:
\begin{equation}
1 + \frac{a^2A^2C_1C_2\gamma^2}{(a^2-l^2)^2+(2al)^2}+\frac{Ab^2BC_3C_4\gamma^2}{(b^2-l^2)^2+(2bl)^2}=0
\label{eq:ImPrtDnmntr}
\end{equation}
\ni Noting the relationship: $C = C_1 = \frac{5}{4}C_2 = 4C_3 = 4C_4$, $C$ can be eliminated by combining equations \ref{eq:RlPrtDnmntr} and \ref{eq:ImPrtDnmntr}, yielding a form for $l$:
\begin{equation}
(a^2-l^2)(\frac{8}{5}a^2A(b^2+l^2)^2)-\frac{bB}{16}(a^2+l^2)^2(b(a^2-l^2)+a(b^2-l^2))=0
\label{eq:EqnForl}
\end{equation}
\ni This can readily be solved by substituting $y = l^2$ and applying the cubic formula. The solution for $l$ can then be substituted back into the form for either the real or imaginary part of the denominator to find C. For example:
\begin{equation}
C^2 = \frac{-(a^2+l^2)^2(b^2+l^2)^2}{0.8(aA\gamma)^2(b^2+l^2)^2-AB(\frac{b\gamma}{4})^2(a^2+l^2)^2}
\end{equation}

\begin{figure*}
\centering
\captionsetup{width=\AppendixFigureWidth}
\includegraphics[width= 0.6666\textwidth]{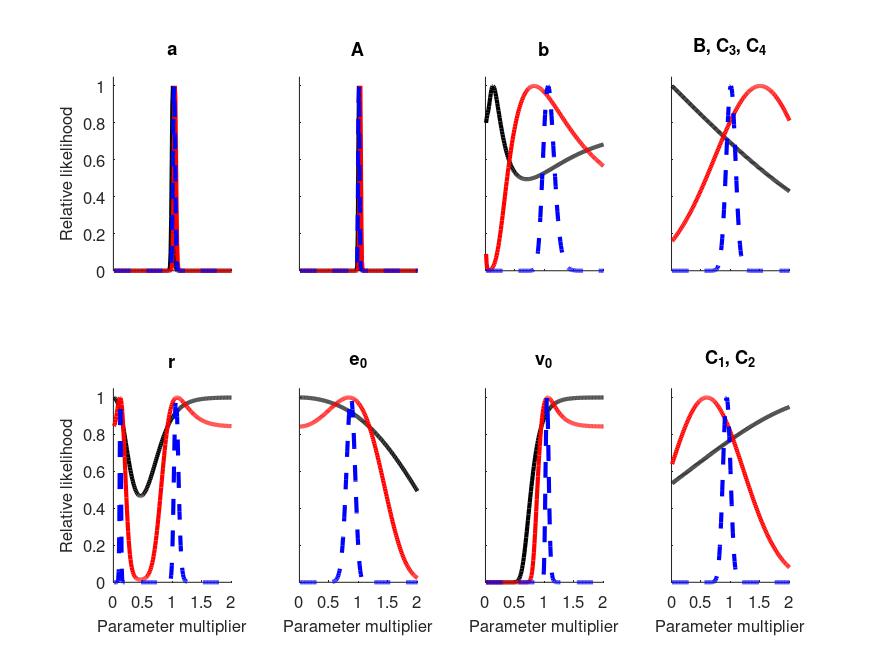}
\caption{Accuracy of the recovery of the parameters of the linearised Jansen-Rit model using DCM. The relative likelihood of parameter values for illustrative values of connectivity ($C = 50$, black; $C = 100$, red; $C = 300$, dashed blue) are shown against the relative parameter value (a parameter multiplier of one corresponds to the true value). The results here are from fitting the transfer function to the Fourier transform of the model output allowing knowledge of the rate of input spikes to the model.}
\label{fig:Figure9Appendix}  
\end{figure*} 

\begin{figure*}
\centering
\captionsetup{width=\AppendixFigureWidth}
\includegraphics[width= 0.6666\textwidth]{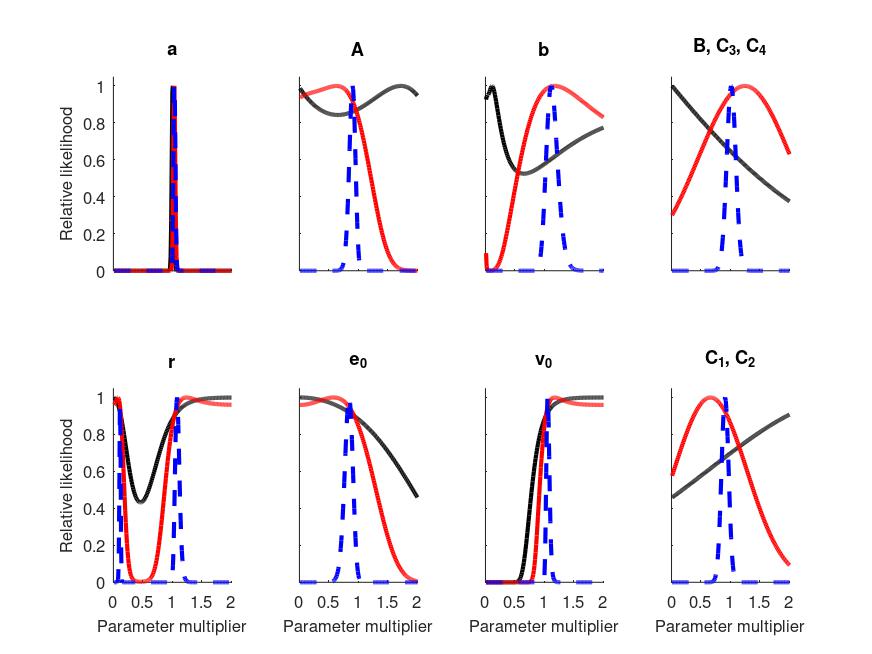}
\caption{Accuracy of the recovery of the parameters of the linearised Jansen-Rit model using DCM. The relative likelihood of parameter values for illustrative values of connectivity ($C = 50$, black; $C = 100$, red; $C = 300$, dashed blue) are shown against the relative parameter value (a parameter multiplier of one corresponds to the true value). The results here are from fitting the transfer function to the Fourier transform of the model output with both the transfer function and Fourier transform normalised, reflecting the scenario where the input spike rate is unknown.}
\label{fig:Figure10Appendix}  
\end{figure*}

\begin{figure*}
\centering
\captionsetup{width=\AppendixFigureWidth}
\includegraphics[width= 0.6666\textwidth]{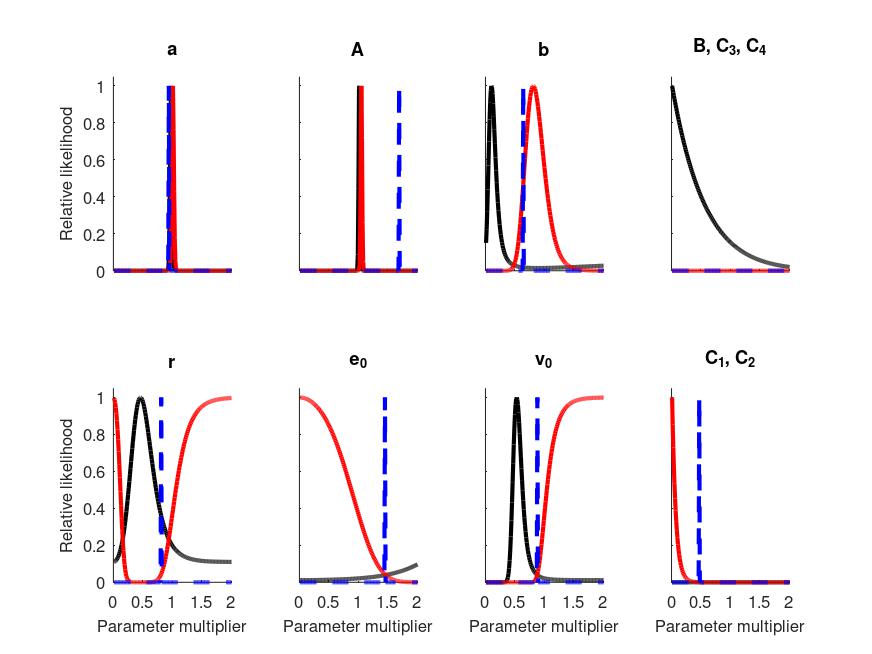}
\caption{Accuracy of the recovery of the parameters of the Jansen-Rit model with the orignal sigmoid function using DCM. The relative likelihood of parameter values for illustrative values of connectivity ($C = 50$, black; $C = 100$, red; $C = 300$, dashed blue) are shown against the relative parameter value (a parameter multiplier of one corresponds to the true value). The results here are from fitting the transfer function to the Fourier transform of the model output allowing knowledge of the rate of input spikes to the model.}
\label{fig:Figure11Appendix}  
\end{figure*}

\begin{figure*}
\centering
\captionsetup{width=\AppendixFigureWidth}
\includegraphics[width= 0.6666\textwidth]{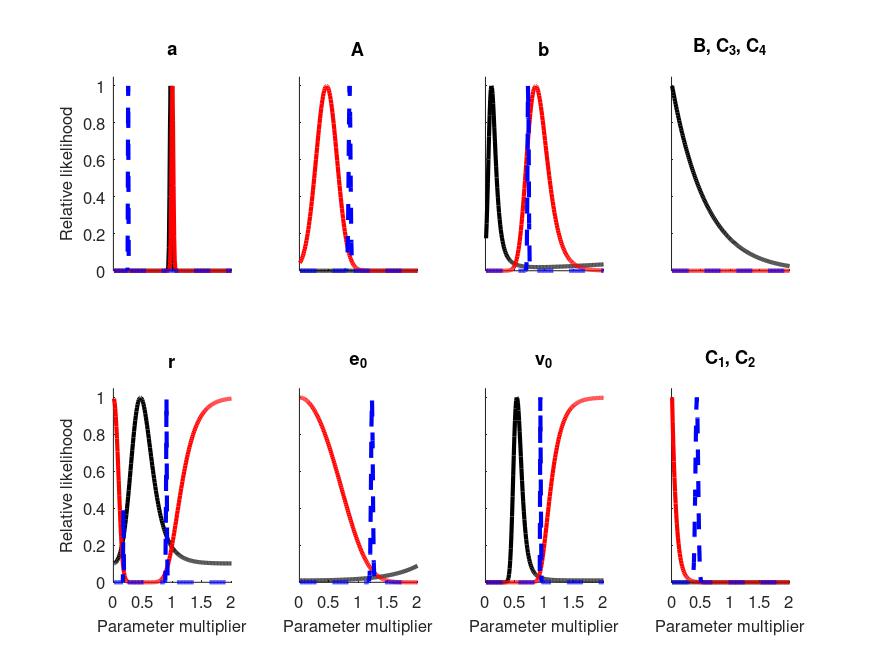}
\caption{Accuracy of the recovery of the parameters of the Jansen-Rit model with the orignal sigmoid function using DCM. The relative likelihood of parameter values for illustrative values of connectivity ($C = 50$, black; $C = 100$, red; $C = 300$, dashed blue) are shown against the relative parameter value (a parameter multiplier of one corresponds to the true value). The results here are from fitting the transfer function to the Fourier transform of the model output with both the transfer function and Fourier transform normalised, reflecting the scenario where the input spike rate is unknown.}
\label{fig:Figure12Appendix}  
\end{figure*}

\begin{figure*}
\centering
\captionsetup{width=\AppendixFigureWidth}
\includegraphics[width= 0.6666\textwidth]{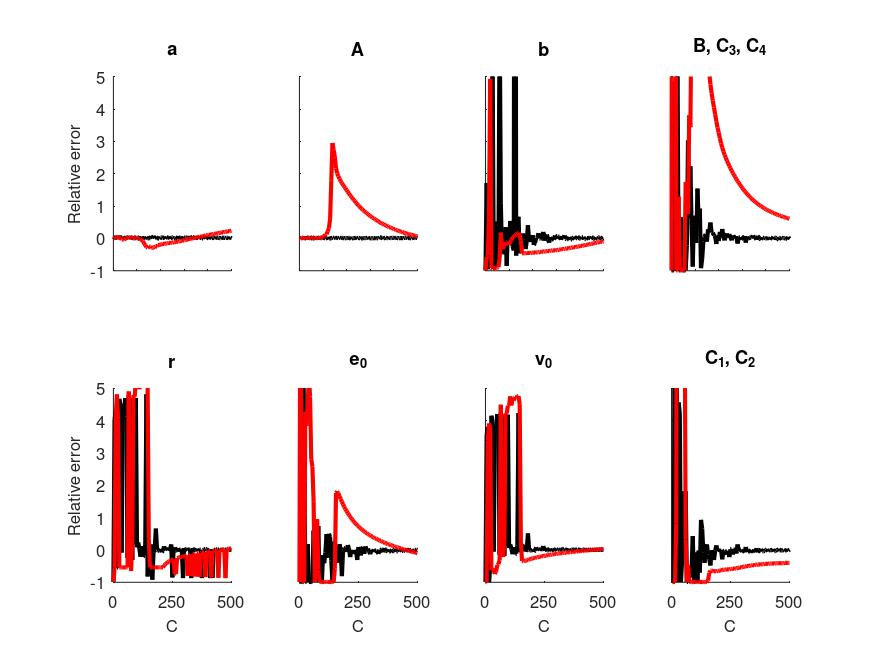}
\caption{Accuracy of the recovery of the parameters of the linearised Jansen-Rit model (black) and original model (red) given knowledge of the input spike rate. The relative error in the maximum likelihood estimates of the parameters is shown as a function of the coupling constant, $C$, used in the model. }
\label{fig:Figure13Appendix}  
\end{figure*}

\begin{figure*}
\centering
\captionsetup{width=\AppendixFigureWidth}
\includegraphics[width= 0.6666\textwidth]{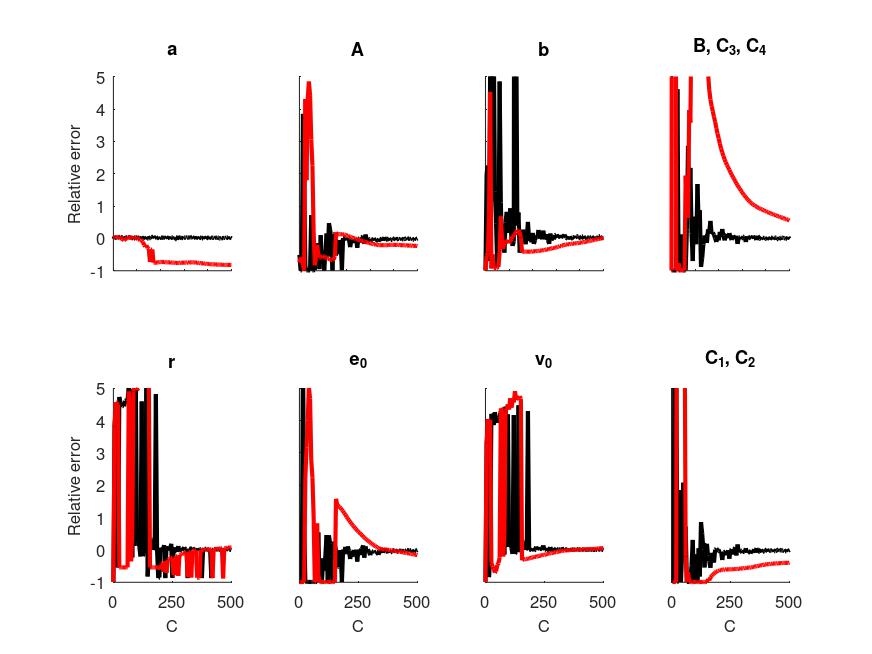}
\caption{Accuracy of the recovery of the parameters of the linearised Jansen-Rit model (black) and original model (red) without knowledge of the input spike rate. To apply DCM both the transfer function and Fourier transform were normalised by their means. The relative error in the maximum likelihood estimates of the parameters is shown as a function of the coupling constant, $C$, used in the model.}
\label{fig:Figure14Appendix}  
\end{figure*}

\begin{figure*}
\centering
\captionsetup{width=\AppendixFigureWidth}
\includegraphics[width= 0.8333\textwidth]{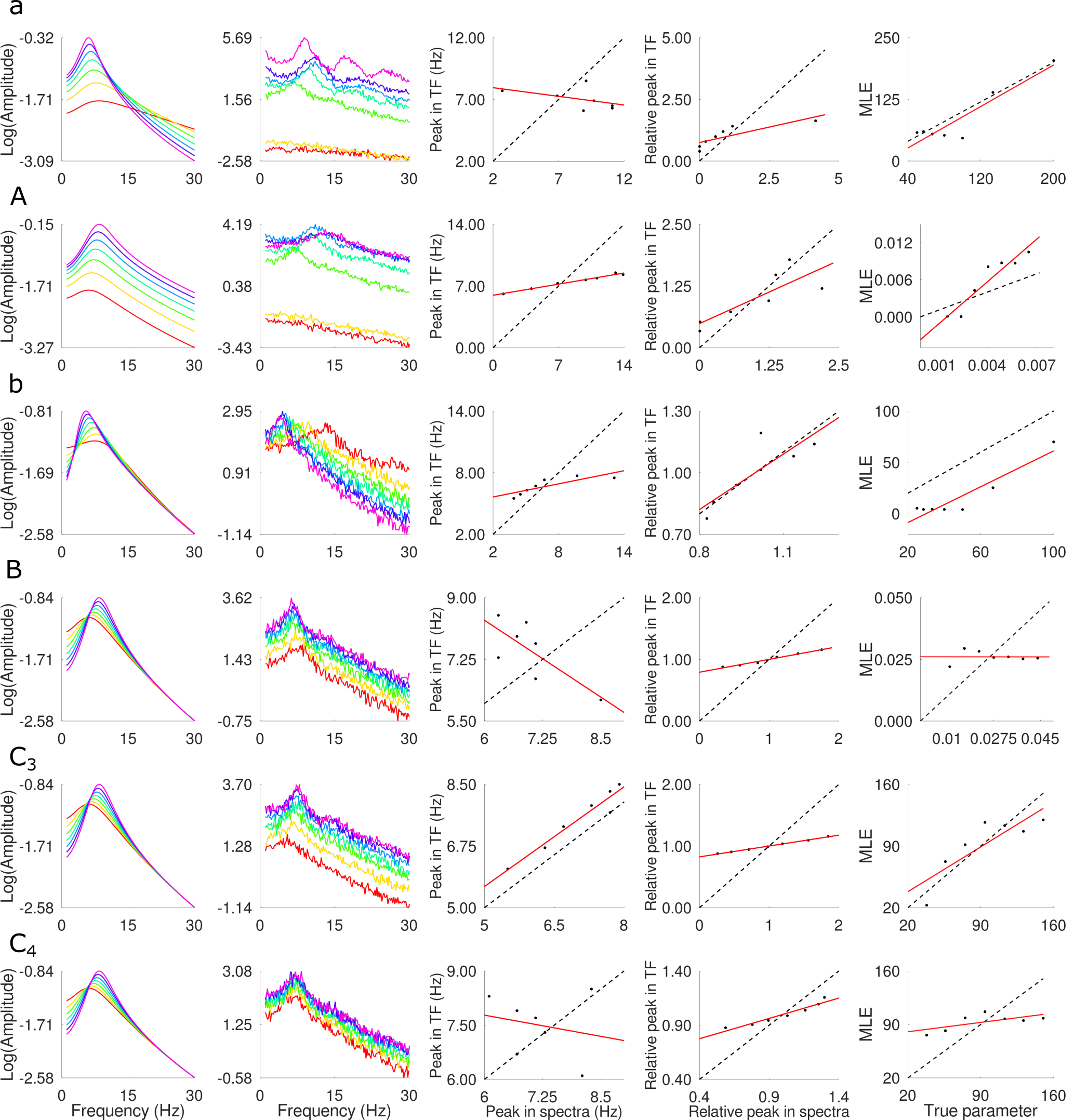}
\caption{Recovery of the parameters of networks of spiking neurons. Each row corresponds to variation of a single parameter as indicated ($a$, $A$, $b$, $B$, $C_3$, and $C_4$). The first column shows the transfer functions for the equivalent linearised Jansen-Rit model as the relevant parameter is varied from 50\% (red) to 200\% (magenta). The second column shows the Fourier transform of the model LFP as the relevant parameter is varied from 50\% (red) to 200\% (magenta). The first and second columns have a logarithmic scale on the y-axis. Subsequent columns show the relationship between the frequency at which the peak in the spectra and the peak in the transfer function occurs, the relative amplitude of the peak in the spectra and the transfer function, and the true parameter and the maximum likelihood estimate. In each case the solid red line shows a least squares fit to the data points and the dashed black line shows the line $y=x$.}
\label{fig:Figure15Appendix}  
\end{figure*}

\end{adjustwidth}


\end{onehalfspace}

\end{document}